\documentclass[onecolumn,showpacs]{revtex4}

\usepackage{graphicx}% Include figure files
\usepackage{dcolumn}% Align table columns on decimal point
\usepackage{bm}% bold math

\input epsf

\begin{document}

\title{\Large Quintessence Problem and Brans-Dicke Theory}

\author{\bf Subenoy Chakraborty}
\email{subenoyc@yahoo.co.in}
\author{\bf N. C.  Chakraborty}
\author{\bf Ujjal Debnath}
\email{ujjaldebnath@yahoo.com}

\affiliation{Inter University Centre for Astronomy and
Astrophysics, Post Bag-4, Ganeshkhind, Pune 411007, India\\
\footnote{Permanent address for correspondence.}Department of
Mathematics, Jadavpur University, Calcutta-32, India.}

\date{\today}

\begin{abstract}
It  has  been  shown  that  Brans-Dicke (BD) theory  in
anisotropic  cosmological model can  alone    solve the
quintessence problem  and  we  have  accelerated  expanding
universe  without  any quintessence  matter. Also the flatness
problem  has  been  discussed  in  this  context.
\end{abstract}


\maketitle

\section{\normalsize\bf{Introduction}}

For  the  last  few  decades, it  is  generally  believed  that
after  a (small) period  of  inflationary era  (when  the
universe  was  accelerated)  the  universe  has  been  evolving
in   a  Big  bang   scenario (deceleration  of  the universe)
and  most  of  the  present day  observations  are  in  accord
with  this  cosmological model (known  as  Standard Cosmological
Model). But difficulty  has  started  when recent  data from high
redshift  supernovae [1,2] (in  fact from  luminosity redshift
relation  of  type I a  supernovas upto  about  $z = 1$) suggest
an  accelerated  universe  at  the present  epoch  (known  as
quintessence  problem). So  there must  be  some matter field
(known  as  Quintessence  matter or  Q-matter) which  is either
neglected  or  unknown, responsible  for this accelerated
universe. From  theoretical point  of  view, a lot  of  works
[3-8] has  been  done  to solve  this quintessence  problem and
possible  candidates for Q-matter are  Cosmological Constant (or
more  generally a variable cosmological  term), a scalar field
[9,10] with  a potential  giving  rise  to  a negative pressure at
the present  epoch, a  dissipative fluid with  an effective
negative stress [11] and  more exotic matter  like  a frustrated
network of  non-abelian cosmic strings  or domain walls [12,13].
In these  works, it  is generally assumed that  Q-matter behaves
as a perfect  fluid with barotropic equation  of state and  most
of them  have considered spatially  flat model  of the universe
(only the work  of Chimento  etal [11] has  been done for  open
model  of the universe). Recently, Banerjee etal [14] have shown
that it is  possible  to  have an accelerated universe  with BD-theory
in  Friedmann  model  without  any matter.\\

In  this  paper  we  have  generalized  their  work  considering
anisotropic  models  of  the  universe. The  paper  is organized
as  follows:  In  section II, we  have  studied field equations
and   with  its  solutions. In  section III, we have discussed
the  conformal  transformations  to  the model and also  solve
the  flatness  problem. In  the  last section (i.e., section IV)
we  give  some  remarks  of  this paper.

\section{\normalsize\bf{Field  equations  and  solutions}}
In  this  paper, we  consider  anisotropic  space-time  model
described  by  the  line  element

\begin{equation}
ds^{2}=-dt^{2}+a^{2}dx^{2}+b^{2}d\Omega_{k}^{2}
\end{equation}

where  $a$  and  $b$  are  functions  of  time  alone : we  note
that

\begin{eqnarray}d\Omega_{k}^{2}= \left\{\begin{array}{lll}
dy^{2}+dz^{2}, ~~~~~~~~~~~~ \text{when} ~~~k=0 ~~~~ ( \text{Bianchi ~I ~model})\\
d\theta^{2}+sin^{2}\theta d\phi^{2}, ~~~~~ \text{when} ~~~k=+1~~
( \text{Kantowaski-Sachs~ model})\\
d\theta^{2}+sinh^{2}\theta d\phi^{2}, ~~~ \text{when} ~~~k=-1 ~~(
\text{Bianchi~ III~ model})\nonumber
\end{array}\right.
\end{eqnarray}

Here  $k$  is  the  curvature  index  of  the  corresponding
2-space, so  that  the  above  three  types  are  described  by
Thorne [15]  as  Euclidean, open  and  semi  closed
respectively.\\

Now, in  BD  theory, assuming  a  perfect  fluid  distribution
as  only  matter  field, the  field  equations  for  the  above
space-time  symmetry  are

\begin{equation}
\frac{\ddot{a}}{a}+2\frac{\ddot{b}}{b}=-\frac{1}{(3+2\omega)\phi}\left[
(2+\omega)\rho_{_{f}}+3(1+\omega)p_{_{f}}\right]-\omega\left(\frac{\dot{\phi}}
{\phi}\right)^{2}-\frac{\ddot{\phi}}{\phi}
\end{equation}

\begin{equation}
\left(\frac{\dot{b}}{b}
\right)^{2}+2\frac{\dot{a}}{a}\frac{\dot{b}}{b}=\frac{\rho_{_{f}}}{\phi}-\frac{k}{b^{2}}-\left(\frac{\dot{a}}{a}+2\frac{\dot{b}}{b}
\right)\frac{\dot{\phi}}{\phi}+\frac{\omega}{2}\left(\frac{\dot{\phi}}{\phi}
\right)^{2}
\end{equation}

and  the  wave  equation  for  BD  scalar  field  is

\begin{equation}
\ddot{\phi}+\left(\frac{\dot{a}}{a}+2\frac{\dot{b}}{b}
\right)\dot{\phi}=\frac{1}{3+2\omega}\left(\rho_{_{f}}-3p_{_{f}}
\right)
\end{equation}

Here  $\rho_{_{f}}$  and  $p_{_{f}}$  are  the  density  and
hydrostatic pressure respectively  of  the  fluid  distribution,
obeying  the barotropic  equation  of  state

$$
p_{_{f}}=(\gamma_{_{f}}-1)\rho_{_{f}}
$$

($\gamma_{_{f}}$ being  the  constant  adiabatic  index  of  the
fluid causality  demands  $0\le \gamma_{_{f}}\le 2$) and  as
usual $\omega$ is the BD coupling parameter. Now  the  above
field equations, via the Bianchi identities  lead  to  the energy
conservation equation

\begin{equation}
\dot{\rho}_{_{f}}+\left(\frac{\dot{a}}{a}+2\frac{\dot{b}}{b}
\right)\left(\rho_{_{f}}+p_{_{f}}\right)=0
\end{equation}

At  present, the  universe  is  considered  as  matter dominated
(i.e., filled  with  cold  matter (dust) of negligible  pressure)
so considering $p_{_{f}}= 0$, we  have  from equation (5) (after
integration)

\begin{equation}
\rho_{_{f}}=\frac{\rho_{_{0}}}{V}
\end{equation}

where $V=ab^{2}$ is the volume index  at  the present instant and
$\rho_{_{0}}$ is  an  integration  constant. Also  using
$p_{_{f}}=0$  and
 equation (6), the  wave equation  has  a  first
integral

\begin{equation}
V\dot{\phi}=\frac{\rho_{_{0}}t}{3+2\omega}+c_{_{0}}
\end{equation}

We  now  assume  a  power-law  form  of  the  scale  factors
keeping  in  mind  that  we  must  have  an  accelerated
universe  to  match  the  recent  observations. So  we  take

\begin{equation}
a(t)=a_{_{0}}t^{\alpha},~~~b(t)=b_{_{0}}t^{\beta}
\end{equation}

and  consequently

\begin{equation}
V=V_{_{0}}t^{\alpha+2\beta},~~(V_{_{0}}=a_{_{0}}b_{_{0}}^{2})
\end{equation}

where  ($a_{_{0}},b_{_{0}}$) are  positive  constants  and
($\alpha,\beta$) are real constants. Thus  the  matter  density
and the BD-scalar field  takes  the  form

\begin{equation}
\rho_{_{f}}=\rho_{_{0}}t^{-(\alpha+2\beta)},~~~(\rho_{_{f}}=\rho_{_{0}}/V_{_{0}})
\end{equation}

and

\begin{equation}
\phi=\frac{\rho_{_{1}}t^{2-\alpha-2\beta}}{(3+2\omega)(2-\alpha-2\beta)}+
\frac{c_{_{0}}t^{1-\alpha-2\beta}}{V_{_{0}}(1-\alpha-2\beta)}
\end{equation}

The  field  equations  will  be  consistent  for  the  above
solutions (eqs.(8)-(11)) provided  we  have  the  following
restrictions  on  the  parameters

\begin{equation}
c_{_{0}}=0,~~\beta=1~~~and
\end{equation}
either
\begin{equation}
\alpha=1,~~\omega=-2\left(1+\frac{k}{3b_{_{0}}^{2}}\right)
\end{equation}
or
\begin{equation}
\omega=-2,~~\frac{k}{b_{_{0}}^{2}}=\alpha-1
\end{equation}

It  is  to  be  noted  that  both  the  cases  coincide  for
flat  model  of  the  universe  and  we  have  the  solution

\begin{equation}
a=a_{_{0}}t,~~b=b_{_{0}}t,~~\rho_{_{f}}=\rho_{_{1}}t^{-3},~~
\phi=\rho_{_{1}}t^{-1},~~2\omega+3=-1
\end{equation}

so  for $k\ne 0$, the  solution  in  first  case  is

\begin{equation}
a=a_{_{0}}t,~~b=b_{_{0}}t,~~\rho_{_{f}}=\rho_{_{1}}t^{-3},~~
\phi=-\frac{\rho_{_{1}}t^{-1}}{2\omega+3}
\end{equation}

and

$$
2\omega+3=-\left(1+\frac{4k}{3b_{_{0}}^{2}}\right),
$$

(for $k<0, b_{_{0}}^{2}>4/3$) while  for  the  second  case the
expression for the geometric  and  physical  parameters  are

\begin{equation}
a=a_{_{0}}t,~~b=b_{_{0}}t,~~\rho_{_{f}}=\rho_{_{1}}t^{-(\alpha+2)},~~
\phi=\frac{\rho_{_{1}}}{\alpha}t^{-\alpha},~~2\omega+3=-1
\end{equation}

and  we  have  a  restriction

$$
\frac{k}{b_{_{0}}^{2}}=\alpha-1
$$

The  deceleration  parameter  has  the  expression

\begin{equation}
q=-\left(\frac{\alpha-1}{\alpha+2}\right),
\end{equation}

(Note  that $\alpha=1$ corresponds  to  the  first  two  case)\\

Thus  we  always  have  an  accelerated  model  of  the  universe
(in  the  third  case  i.e., eq.(17)) (except  for $-2\le \alpha
\le 1 $) as predicted  by  recent  Supernova  observation. Hence
for the solutions  represented  by  equations  (15)  and  (16)
$q=0$ i.e., at  present  the  universe  is  in  a  state  of
uniform expansion  and  the  conclusion  is  identical  to  that
of Banerjee  etal [14]. Further, the  solution  corresponds  to
equation  (17)  is  new  as  we  have  a  negative  deceleration
parameter. This  solution  is  valid  for  closed  (or  open)
model  of  the  universe  for  $\alpha>1$ (or $\alpha<1$).
Therefore, it  is possible  to  have  an  accelerated  universe
today  by considering  anisotropy  model  of  the  universe  and
hence the  quintessence  problem  may  be  solved  for  closed,
open or  flat  type  of  model  of  the  universe.

\section{\normalsize\bf{Conformal  transformations : Flatness  problem }}
In  cosmology, the  technique  of  conformal transformation is
often  used  (for  mathematical simplification)  to transform a
non-minimally  coupled  scalar field  to  a minimally coupled
one  [16]. Usually, in  `Jordan Conformal frame' the scalar
field  (also  BD  scalar  field) couples non-minimally to  the
`Einstein  frame'  in  which  the transformed  scalar field  is
minimally  coupled. In the  last section, we  have  developed
the  BD  theory  in Jordan  frame and  to  introduce  the
Einstein  frame  we make the  following transformations:

\begin{equation}
d\eta=\sqrt{\phi}~dt, \bar{a}=\sqrt{\phi}~a,
\bar{b}=\sqrt{\phi}~b, \psi=\text{ln}~\phi,
\bar{\rho}_{_{f}}=\phi^{-2}\rho_{_{f}},
\bar{p}_{_{f}}=\phi^{-2}p_{_{f}}
\end{equation}

As  a  result  the  field  equations  (2)-(4) transformed  to

\begin{equation}
\frac{\bar{a}''}{\bar{a}}+2\frac{\bar{b}''}{\bar{b}}=-\frac{1}{2}\left(\bar{\rho}_{_{f}}+
3\bar{p}_{_{f}}\right)-\frac{(3+2\omega)}{2}\psi'^{2}
\end{equation}

\begin{equation}
\left(\frac{\bar{b}'}{\bar{b}}\right)^{2}+2\frac{\bar{a}'}{\bar{a}}\frac{\bar{b}'}{\bar{b}}+
\frac{k}{\bar{b}^{^{2}}}=\bar{\rho}_{_{f}}+\frac{(3+2\omega)}{4}\psi'^{2}
\end{equation}

and

\begin{equation}
\psi''+\left(\frac{\bar{a}'}{\bar{a}}+2\frac{\bar{b}'}{\bar{b}}\right)\psi'=\frac{1}
{3+2\omega}\left(\bar{\rho}_{_{f}}-3\bar{p}_{_{f}}\right)
\end{equation}

where~~  $' \equiv \frac{d}{d\eta}$.\\

These  equations  are  the  well  known  fields  equations  for
the  anisotropic  cosmological  models (describe here)  with a
minimally  coupled  scalar  field  $\psi$ (massless). This scalar
field  behaves  like  a  `stiff'  perfect  fluid  with equation of
state

\begin{equation}
\bar{p}_{_{\psi}}=\bar{\rho}_{_{\psi}}=\frac{\psi'^{2}}{16\pi G}
\end{equation}

In  Einstein  frame, the  total  stress-energy  tensor  is
conserved, but  the  scalar  field  and  normal  matter  change
energy  according  to

\begin{equation}
\bar{\rho}_{_{f}}+\left(\frac{\bar{a}'}{\bar{a}}+2\frac{\bar{b}'}{\bar{b}}\right)
\left(\bar{\rho}_{_{f}}+\bar{p}_{_{f}}\right)=-\left[\bar{\rho}_{_{\psi}}+
\left(\frac{\bar{a}'}{\bar{a}}+2\frac{\bar{b}'}{\bar{b}}\right)\left(\bar{\rho}_{_{\psi}}+
\bar{p}_{_{\psi}}\right)\right]=-\frac{\psi'}{2}\left(\bar{\rho}_{_{f}}-
3\bar{p}_{_{f}}\right)
\end{equation}

Thus  combining  the  two  energy  densities  we  have  from
the  above  equation

\begin{equation}
\bar{\rho}'+3\gamma H\bar{\rho}=0
\end{equation}

Here
$H=\frac{1}{3}\left(\frac{\bar{a}'}{\bar{a}}+2\frac{\bar{b}'}{\bar{b}}\right)$
is  the  Hubble  parameter  in  the Einstein frame and  $\gamma$
is the  average  barotropic  index  defined  as

\begin{equation}
\gamma\Omega=\gamma_{_{f}}\Omega_{_{f}}+\gamma_{_{\psi}}\Omega_{_{\psi}}
\end{equation}

where

\begin{equation}
\Omega=\Omega_{_{f}}+\Omega_{_{\psi}}=\frac{\bar{\rho}}{3H^{^{2}}}
\end{equation}

is  the density  parameter.\\

From  equations  (21)  and  (25)  after  some  algebra, we  have

\begin{equation}
\Omega'=\Omega(\Omega-1)[\gamma H_{_{a}}+2(\gamma-1)H_{_{b}}]
\end{equation}

where ~~ $H_{_{a}}=\frac{\bar{a}'}{\bar{a}}$    and
$H_{_{b}}=\frac{\bar{b}'}{\bar{b}}$ .\\

This evolution in $\Omega$ shows that  $\Omega=1$ is a possible
solution  of it and  for stability of  this solution, we  have

\begin{equation}
\gamma<\frac{2}{3},
\end{equation}

for  the  solutions  given  in  equations  (15)  and  (16)  and
the  restriction  is

\begin{equation}
\gamma<\frac{2}{\alpha+3},
\end{equation}

for  the  solution  (17).\\

Since  the  adiabatic  indices  do  not  change  due  to
conformal  transformation  so  we  take  $\gamma_{_{f}}=1$ (since
$p_{_{f}}=0$) and $\gamma_{_{\psi}}=2$. Hence  from  (26)  and
(27), we have

\begin{equation}
\gamma=\frac{\Omega_{_{f}}+2\Omega_{_{\psi}}}{\Omega_{_{f}}+\Omega_{_{\psi}}}
\end{equation}

Now  due  to  upper  limit  of  $\gamma$ (given above) we  must
have the inequalities

$$
\Omega_{_{f}}<4|\Omega_{_{\psi}}|
$$
$$
OR
$$

\begin{equation}
\Omega_{_{f}}<\frac{2(\alpha+1)}{\alpha}|\Omega_{_{\psi}}|
\end{equation}

according  as  $\gamma$ is  restricted  by  (29)  or  (30). Also
from the  field  equation  (21), the  curvature  parameter
$\Omega_{_{k}}=-k/\bar{b}^{2}$ vanishes for  the  solution
$\Omega=1$, provided  we are restricted to $\alpha=1$. Therefore,
depending  on the relative magnitude  of the energies  of matter
and  the BD-scalar  field (as  in  isotropic  case) it  is
possible  to  have  a stable  solution  corresponding  to
$\Omega=1$ and hence  the flatness  problem  can  be  solved  to
extent.

\section{\normalsize\bf{Concluding  remarks }}

In  this  work, we  have  considered  three  anisotropic
cosmological  models  namely, Bianchi III ($k<0$), axially
symmetric  Bianchi I ($k=0$)  and  Kantowski-Sachs ($k>0$)
space-time. We  have shown  that  the  anisotropic character is
responsible  for  getting  an  accelerated  model of  the
universe. In  fact, for  the  present  model  the  shear scalar
is given  by $\sigma^{2}=\frac{2}{3}(\alpha-1)^{2}$, so  the
deceleration parameter (see eq. (18)) is  proportional  to
$\sqrt{\sigma}$ and it shows  how anisotropy characterizes  the
accelerating  or decelerating universe. In other  words  we can
say  that anisotropic nature  of the universe  has  an effect on
the quintessence problem. The problem  for negative coupling
constant $\omega$ is same  as in isotropic  case and  there is
problem in big-bang nucleosynthesis  scenario as  claimed  by
Banerjee etal [14]. Lastly, in  this  case the  modified version
of BD-theory (where  the  coupling parameter $\omega$ is a
function of the scalar  field) is  very similar  to  that for
isotropic case, so  we  have  not presented  here.\\

{\bf Acknowledgement:}\\

One  of  the  author (S.C.)  is  thankful  to  IUCAA  (where  a
part  of  the  work  has been  done) for  worm  hospitality and
facilities  for  this  work. Also  U.D. is thankful  to C.S.I.R.,
Govt. of  India  for  awarding  a  Junior  Research  Fellowship.\\

{\bf References:}\\
\\
$[1]$  Perlmutter  S  etal  {\it Nature} (London)  {\bf 391}  51
(1998);  {\it Astrophys.  J.}  {\bf 517}  565 (1999).\\
$[2]$  Riess A G etal  {\it Astrophys. J.}  {\bf 116}  1009
(1998); Garnavich P M etal  {\it Astrophys. J.}  {\bf 509} 74
(1998).\\
$[3]$  Ostriker  J P and Steinhardt P J  {\it Nature} (London)
{\bf 377} 600 (1995).\\
$[4]$  Peebles J P  E  {\it Astrophys.  J.} {\bf 284} 439
(1984).\\
$[5]$  Wang L, Caldwell  R, Ostriker J P, Ateinhart  P  J {\it
Astrophys. J.} {\bf 530}  17 (2000).\\
$[6]$  Caldwell R R, Dave  R and Steinhardt P  J  {\it Phys. Rev.
Lett.} {\bf 80} 1582 (1998).\\
$[7]$  Perlmutter S, Turner M  S and White M  {\it Phys. Rev.
Lett.} {\bf 83}  670 (1999); Dodelson  S, Kaplinghat  M  and
Stewart  E  ``Tracking  Oscillating  energy'' {\it
astro-ph}/0002360.\\
$[8]$  Faraoni  V {\it Phys.  Rev.  D} {\bf 62} 023504 (2000).\\
$[9]$  The scalar field  was  introduced  by  Peebles and Ratra:\\
Peebles  J  P  E  and  B  Ratra  {\it Astrophys.  J.  Lett.} {\bf
325} L17 (1988); B  Ratra  and  Peebles  J  P  E  {\it Phys. Rev.
D} {\bf 37} 3406 (1988).\\
$[10]$  For  recent  work:\\ Ott, T. {\it Phys. Rev. D} {\bf 64}
023518 (2001); Hwang, J.-c. and  H. Noh,  {\it Phys. Rev. D} {\bf
64} 103509 (2001);  Ferreira, P. G. and  M. Joyce,  {\it Phys.
Rev. D} {\bf 58}  023503 (1998).\\ For  a  review  see,  Peebles
J  P  E  and  B  Ratra,  {\it astro-ph}/0207347.\\
$[11]$  Chimento L P, Jakubi  A  S  and  Pavon  D  {\it Phys.
Rev.
D} {\bf 62} 063508 (2000).\\
$[12]$  Bucher  M  and  Spergel D  {\it Phys. Rev. D} {\bf 60}
043505 (1999).\\
$[13]$  Battye  R  A, Bucher M and Spergel  D  ``Domain wall
dominated universe'', {\it astro-ph}/990847.\\
$[14]$  Banerjee  N and Povon  D  {\it Phys. Rev.  D}  {\bf 63}
043504
(2001).\\
$[15]$  Thorne  K  S  {\it Astrophys.  J.} {\bf 148} 51 (1967).\\
$[16]$  Faraoni V, Gunzig  E  and Nardone P   {\it Fundam. Cosm.
Phys.} {\bf 20}  121 (1999).\\

\end{document}